\documentclass[iop]{emulateapj}
\usepackage{rotating}
\usepackage{epsfig}

\begin{document}

\slugcomment{ApJ Letters, in press}

\title{Pre- and Post-burst Radio Observations of the\\ Class 0 Protostar HOPS 383 in Orion}

\author{
Roberto Galv\'an-Madrid\altaffilmark{1}, Luis F. Rodr{\'\i}guez\altaffilmark{1,2}, Hauyu Baobab Liu\altaffilmark{3} \\
Gr\'ainne Costigan\altaffilmark{4}, Aina Palau\altaffilmark{1},  Luis A. Zapata\altaffilmark{1}, 
Laurent Loinard\altaffilmark{1}
} 

\altaffiltext{1}{Centro de Radioastronom\'\i a y Astrof\'\i sica, 
UNAM, Apdo. Postal 3-72 (Xangari), 58089 Morelia, Michoac\'an, M\'exico}
\altaffiltext{2}{Astronomy Department, Faculty of Science, King Abdulaziz University, 
P.O. Box 80203, Jeddah 21589, Saudi Arabia}
\altaffiltext{3}{Academia Sinica, Institute of Astronomy and Astrophysics, PO Box 23-141, Taipei, 106, Taiwan}
\altaffiltext{4}{Leiden Observatory, University of Leiden, PB 9513, 2300 RA, Leiden, The Netherlands}

\email{r.galvan@crya.unam.mx}

\begin{abstract}

There is increasing evidence that episodic accretion is a common phenomenon in Young Stellar Objects (YSOs). 
Recently, the source HOPS 383 in Orion was reported to have a $\times 35$ mid-infrared -- and bolometric -- luminosity 
increase between 2004 and 
2008, constituting the first clear example of a class 0 YSO (a protostar) with a large accretion burst.  
The usual assumption that in YSOs accretion and ejection follow each other in time needs to be tested. 
Radio jets at centimeter wavelengths are often the only way of tracing the jets from embedded protostars. 
We searched the Very Large Array archive for the available observations of the radio counterpart of HOPS 383. 
The data show that the radio flux of HOPS 383 varies only mildly from January 1998 to December 2014, 
staying at the 
level of $\sim 200$ to 300 $\mu$Jy in the X band ($\sim 9$ GHz), with a typical uncertainty of 10 to 20 $\mu$Jy 
in each measurement. 
We interpret the absence of a radio burst as suggesting that accretion and ejection enhancements do not 
follow each other in time, at least not within timescales shorter than a few years. Time monitoring of more
objects and specific predictions from simulations are needed to clarify the details of the connection 
betwen accretion and jets/winds in YSOs. 

\end{abstract}  

\keywords{
stars: formation --
stars: protostars --
stars: radio continuum --
ISM: jets and outflows
}

\section{Introduction}

The physical mechanisms of gas accretion and ejection are fundamental aspects in the 
paradigm of low-mass star formation. In its most commonly assumed form, this paradigm requires the 
quasi-stationary evolution of the forming stars \citep[e.g.,][]{Shu94}.
This assumption is seriously challenged by the observational 
evidence of episodic accretion. First, a handful of individual bursting objects such as FU Ori and EX Lup provide 
evidence that large accretion bursts in Young Stellar Objects (YSOs) do actually happen, with durations from a few months  
to $\sim 100$ yr  
\citep[e.g.,][]{HK96,Herbig08,Audard14}. Second, surveys show that YSOs are systematically underluminous with 
respect to the expectation of steady-accretion models \citep[e.g.,][]{Evans09,Kryukova12}. Episodes of  
significantly increased accretion would alleviate this discrepancy \citep[e.g.,][]{Voro09,Zhu09}. 
Third, episodic accretion in protostars 
-- or class 0 and I YSOs --- has also been invoked to explain the observed 
spread of the Herzprung-Russell diagram of young star clusters \citep{Baraffe12}. 

Although time monitoring campaigns are revealing a large diversity of variation phenomena in YSOs 
\citep[e.g.,][]{Costigan14, Guenter14},  
it is still not clear whether or not accretion bursts are a widespread phenomenon in YSO evolution. The evidence 
is particularly scarce for the case of protostars. Only a few class I or young class II objects have been reported 
to show infrared bursts related to accretion \citep[e.g.,][]{RA04,Caratti11,Covey11,Fischer12,Muzerolle13}. 

Recently, \cite{Safron15} reported an impressive outburst in the embedded object HOPS 383 in Orion. 
This object was classified as a protostar in the {\it Spitzer} survey presented in \cite{Megeath12} and 
in the {\it Herschel}+APEX survey presented in \cite{Stutz13}. 
Based on the ratio of submm-to-bolometric luminosity and in the characteristic 
temperature of the spectral energy distribution, 
\cite{Safron15} conclude that the object is a class 0 YSO, making HOPS 383 the youngest protostar 
known with an accretion-related burst.  
The outburst is most notably seen at a wavelength of $24~\mu$m, in which the brightness increased 
by a factor $\times 35$ from 2004 to 2008. From their extensive IR to submm follow up, \cite{Safron15} 
find no evidence for significant fading of HOPS 383 up through 2012.

Radio observations are a key tool to study YSOs: the thermal free-free emitting radio jets are often the only 
way to probe the bipolar ejection of material close to the protostar in the optically-obscured class 0 objects 
\citep[e.g.,][]{Rodriguez97, Anglada98}. Radio observations are also useful to probe non-thermal (gyro)synchrotron 
radiation from active YSO magnetospheres \citep[e.g.,][]{Guedel02,Forbrich07}. Systematic studies of YSOs 
show that emission at $\sim 3.5$ cm and its variability in class 0 and I protostars is dominated by 
the free-free radio jets, while in more evolved class II and III YSOs  this emission is dominated 
by the magnetospheric (gyro)synchrotron emission \citep{Dzib13,Liu14}, with possible contributions from 
disk photoevaporation \citep{GM14} or weak radio jets \citep{Rodriguez14}. 

In the standard model of star formation, the magnetohydrodynamical (MHD) launching of jets 
provides for part of the necessary release of specific angular momentum. 
The amount of ejected material is usually taken to be a constant fraction of the accreted material 
\citep[e.g.,][]{Pudritz07,Shang07}. 
There is some observational evidence for this assumption to hold in samples of relatively massive, optically visible YSOs 
\citep[][and references therein]{Frank14}, but it 
has yet to be tested in the most embedded objects and in time domain.  
A good way of doing this test in embedded class 0 objects is to look for possible correlations between the 
free-free emission from radio jets and accretion signatures as seen in the near to mid infrared.  
Bursting protostars offer a unique opportunity to perform this test under different conditions in a given object.  

In this Letter, we report on pre- to post-burst centimeter wavelength observations of the bursting class 0 
YSO HOPS 383. 

\section{Data sets and basic results}

To analyze the time behavior of HOPS 383 at centimeter wavelengths, we looked for the 
available observations in the Very Large Array\footnote{The National Radio Astronomy 
Observatory is operated by Associated Universities, Inc. under cooperative agreement with the
National Science Foundation.} archive. In none of these observations is HOPS 383 at
the phase center, however, it falls inside the field of view,
allowing the determination of its parameters after correction for the
primary beam response. In the following subsections we describe the observation epochs. 
Table 1 summarizes the relevant properties of the data. 

In all the epochs the position 
of the radio source matches within 0.3" to 1.7" with the infrared position reported by 
\cite{Safron15}: $\alpha(\mathrm{J2000})=5^h35^m29\rlap.^{s}81$, $\delta(\mathrm{J2000})=-4^\circ 59' 51\rlap.{''}1$. 
This is a fraction of the synthesized beam in all the epochs in which the radio source appears unresolved (see Table 1), and about 
the synthesized beam size for the 2011 epoch with subarcsecond angular resolution\footnote{The distance to the IR position in this epoch 
is 0.5" to the centroid of the radio emission, and 0.3" to the brightest SE peak.}. Since the next closest radio detection 
is $\approx 120"$
away, and the {\it Spitzer}-MIPS angular resolution is $\sim 6"$, we conclude that the radio detection here reported is 
the counterpart of the bursting protostar reported by \cite{Safron15}.

\begin{deluxetable*}{ccccccc}[h] 
\tabletypesize{\scriptsize} 
\tablecolumns{7} 
\tablewidth{0pt}
\tablecaption{Data Summary \label{tab:data}} 
\tablehead{
\colhead{Epoch} & \colhead{Frequency} & \colhead{Array, HPBW, PA} & \colhead{Phase Center} & \colhead{HOPS383 offset} 
& \colhead{rms noise\footnotemark[1]} & \colhead{Flux Density\footnotemark[2]} \\
\colhead{} & \colhead{[GHz]} & \colhead{[; arcsec$\times$arcsec; deg]} & \colhead{[h:m:s;  deg:arcmin:arcsec]}
 & \colhead{[arcsec]} & \colhead{[$\mu$Jy beam$^{-1}$]} & \colhead{[$\mu$Jy]}
}
\startdata
15.01.1998 & 8.46 & D; $9.2\times7.5$, $-17.8$ & 05:35:24.395; $-05$:01:07.27 & 111.1 & 42 & $230\pm50$ \\ 
14.03.2008+01.06.2008 & 4.86 & DnC+C; $6.0\times4.2$, $49.8$ &  05:35:24.400; $-05$:01:43.25 & 138.2 & 56 & $<224$ \\
14.03.2008+01.06.2008 & 8.46 & DnC+C; $4.8\times3.1$, $57.4$ &  05:35:24.413; $-05$:00:07.25 & 82.2 & 47 & $<188$ \\
13.08.2011 & 4.87 & A; $0.64\times0.42$, $-4.6$ & 05:35:23.420; $-05$:01:30.52 & 137.8 & 10 & $270\pm20$ \\
13.08.2011 & 7.42 & A; $0.44\times0.27$, $-3.6$ & 05:35:23.420; $-05$:01:30.52 & 137.8 & 11 & $310\pm20$ \\
27.10.2014 & 10.0 & C; $2.18\times1.66$, $-1.9$ & 05:35:23.480; $-05$:01:32.30 & 138.5 & 13 & $222\pm12$ \\
03.11.2014 & 10.0 & C; $2.04\times1.81$, $-7.2$ & 05:35:23.480; $-05$:01:32.30 & 138.5 & 15 & $267\pm16$ \\
06.12.2014 & 10.0 & C; $2.29\times1.70$, $-28.3$ & 05:35:23.480; $-05$:01:32.30 & 138.5 & 11 & $293\pm13$ \\
13.12.2014 & 10.0 & C; $2.35\times1.61$, $-11.1$ & 05:35:23.480; $-05$:01:32.30 & 138.5 & 13 & $289\pm15$ \\
15.12.2014 & 10.0 & C; $2.28\times1.65$, $-29.3$ & 05:35:23.480; $-05$:01:32.30 & 138.5 & 10 & $246\pm10$ \\
All 2014 & 9.81	& C, $2.32\times1.62$, $-14.1$ & 05:35:23.480; $-05$:01:32.30 & 138.5 & 6 & $260\pm6$
\enddata
\footnotetext[1]{
rms noise corrected for primary-beam attenuation around the position of HOPS 383: 
$\alpha=5^h35^m29\rlap.^{s}81$, $\delta=-4^\circ 59' 51\rlap.{''}1$. 
\citep{Safron15}.
}
\footnotetext[2]{
Flux density of HOPS 383 corrected for primary-beam attenuation. The errors are only statistical, resulting from 
the noise in the maps and the quality of the fit. The absolute uncertainty of the (J)VLA flux scale is a few percent 
\citep{PB13}. Upper limits in the 2008 epoch are $4\sigma$. 
}
\end{deluxetable*}

\subsection{1998}

These observations were made in 1998 January 13 as part of the project
AR387. The analysis  has been presented and discussed in
\cite{Reipurth99}. The data were obtained at a frequency of 8.46 GHz
in the D configuration. 
\cite{Reipurth99} do not report a radio counterpart to HOPS 383, most probably
because they adopted a stringent limit of 5-$\sigma$ to consider 
a detection. Our reanalysis of these data shows a faint source with
flux density of $0.23\pm0.05$ mJy that coincides spatially with HOPS 383.

\subsection{2008}

These observations were made in 2008 March 14 and June 1 in the C
and DnC configurations at 4.86 and 8.46 GHz as part of project AT359. 
The data from the two epochs were concatenated for increased 
sensitivy, but a counterpart to HOPS 383 was not detected with a
4-$\sigma$ upper limit of 0.22 and 0.19 mJy at 4.86 and 8.46 GHz,
respectively.

\subsection{2011}

These observations were made in 2011 August 13 with the upgraded Jansky Very Large Array 
(JVLA) in the A configuration under project 11A-220. The tuning is 
centered at 4.81 and 7.42 GHz, with total bandwidths of approximately 1 GHz.
This is the only epoch with subarcsecond angular resolution. The source 
associated with HOPS 383 is for the first time resolved into two components (SE and NW) 
with a separation of $\approx 0\rlap.{''}45$ and joined by a ridge of faint emission.    
In maps done with $(u,v)$ weighting close to natural, we measure flux densities 
that are consistent with rest of the detections at other epochs: 
0.27$\pm$0.02 at 4.81 GHz and 0.31$\pm$0.02 mJy at 7.36 GHz (see Table 1). 
The spectral index between these frequencies is 0.3, indicating free-free emission with low 
optical depth. 
Figure 1 shows a map at 7.36 GHz done with weighting between natural and uniform 
to maximize angular resolution. 
The double nature of the radio source is clearly seen, probably from a bipolar radio jet or a binary. 
This map with the highest angular resolution already suffers from significant ($\sim 50\%$) 
flux filtering of extended emission.  

\subsection{2014}

The five more recent observations were made with the upgraded JVLA centered at a rest
frequency of 9.81 GHz (3.0 cm) during 2014 October-December.
These observations are part of project 14B-230.
At that time the array was in its C configuration. 
The total bandwidth of the continuum observations was about 4.0 GHz. 
The individual flux densities determined for the five epochs are listed in Table 1, as well as the 
flux density from the concatenated data. 
There appear to be marginal flux variations within 2014. The maximal flux density is $293\pm13$ $\mu$Jy on 
December 12 2014, which is $71\pm18$ $\mu$Jy above the minimal value of $222\pm12$ $\mu$Jy 
on October 27 2014. Then, without considering the few $\%$ uncertainty in the absolute flux scale \citep{PB13}, 
the largest flux difference within 2014 is significant at $4\sigma$. 
We also tested for intra-day variability by splitting each of the five observing epochs into two time chunks. 
In four of the five epochs, the flux measured in each halve of the data is within 1$\sigma$
of the flux reported in Table 1. In the November 3 2014 epoch, the fluxes in the halved data are within 2$\sigma$ 
of the epoch average, and this is the noisiest of the 2014 observations.
We discard significant intra-day variability. These results are   
consistent with the relatively long variation timescale expected for the radio emission 
of class 0 YSOs, compared to class II and class III \citep{Liu14}. 
Finally, the high sensitivity of these JVLA observations allowed the data to be split into smaller frequency chunks.  
We made images from the concatenated data in four windows, each  
1-GHz wide. The intra-band spectral index of HOPS 383 is consistent with 0, indicating optically-thin free-free emission.

\begin{figure}[!ht]
\centering
\includegraphics[width=\columnwidth]{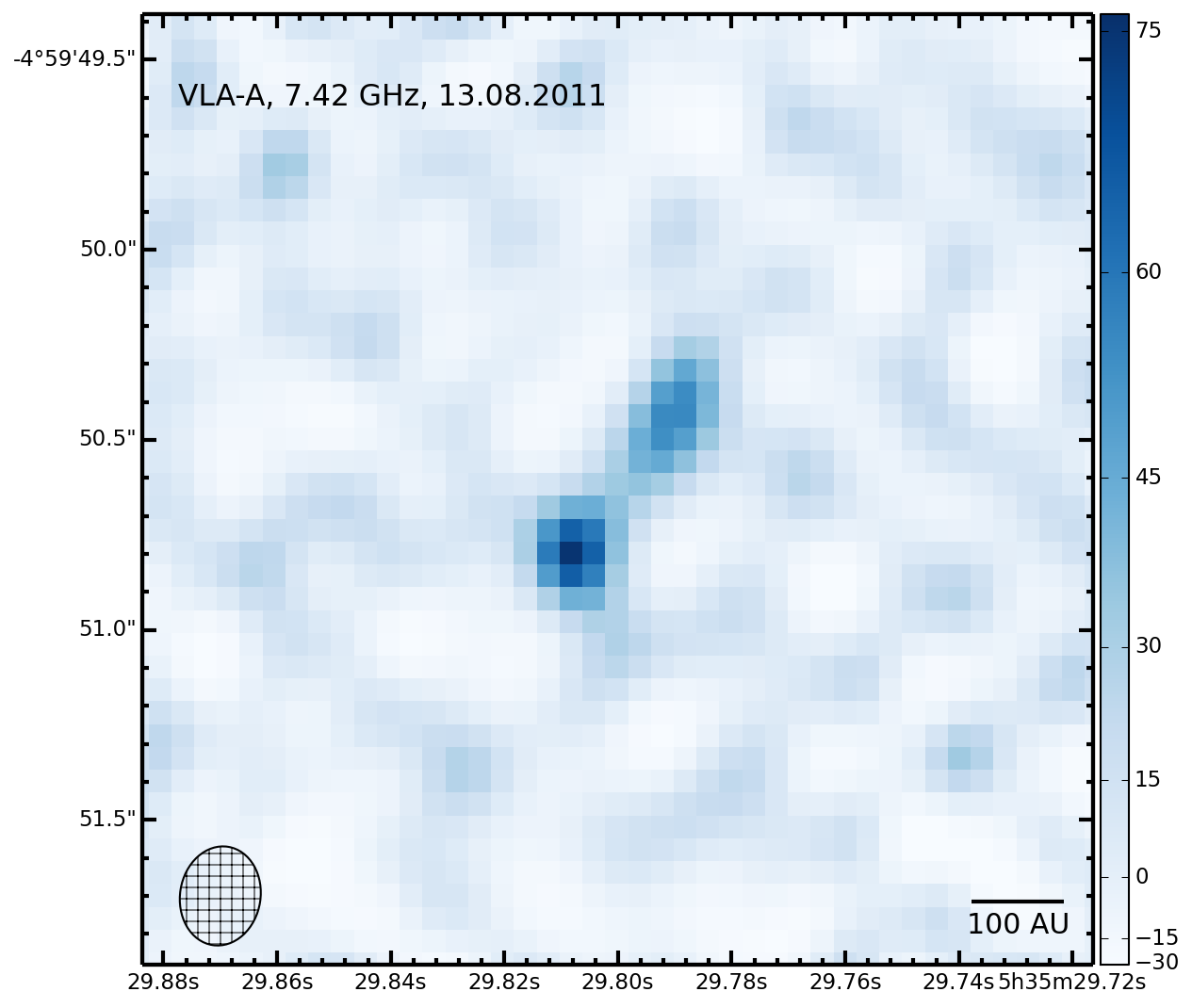}
\caption{\small High-angular resolution JVLA image of the radio counterpart of HOPS 383, observed on August 13, 
2011, and centered at 7.42 GHz (4.0 cm). HPBW$=0\rlap.{''}26 \times 0\rlap.{''}21$, P.A.$=-9.0^\circ$.  The target 
clearly has two components. The peak insensities of the SE and NW components are 75 $\mu$Jy beam$^{-1}$ and 
55 $\mu$Jy beam$^{-1}$, respectively. The rms noise is 11 $\mu$Jy beam$^{-1}$. 
}
\label{fig1}
\end{figure}

\section{Discussion}

\subsection{No radio counterpart to the infrared burst.}

MHD-launched jets that load with them a fraction of the accreted material  
are a basic feature of star formation models \citep[e.g.,][]{Pudritz07,Shang07}. 
Furthermore, there is increasing evidence that this is the case even for the youngest systems:  
the class 0 protostars \citep[e.g.,][]{Li14PPVI}. In these deeply embedded objects, free-free emission 
from radio jets is one of the best ways to trace the outflowing material in the inner few hundred AU 
\citep{Rodriguez97,Anglada98}. The origin of the radio emission is often 
considered to be shock-ionized gas \citep{Anglada96}, but models that include X-ray ionization 
within the X-wind scenario also reproduce the observed properties of radio jets \citep{Shang04}.
Regardless of the detailed jet emission mechanism, in the models it is assumed that an 
increase of accretion is followed by an ejection enhancement. Therefore, if the mid-IR burst truly traces 
a burst of accretion \citep[][]{Safron15}, then a significant increase in the radio flux should follow. 

We find that in HOPS 383 the infrared burst does not have a counterpart in the radio. 
Figure 2 shows the radio light curves. The long-term (decadal) radio flux densities 
show mild variations, staying at a level of $\sim 200$ to 300 $\mu$Jy. 
The earliest 1998 flux density is the same within $1\sigma$ with the latest 2014 measurements (Table 1). 
The average flux density of the detections in the X band (8 to 10 GHz) is $S_\mathrm{X} = 265$ $\mu$Jy, 
with a dispersion of 28 $\mu$Jy, or about the noise in single epochs. 
Taken at face value, it may even seem that during the IR-burst the radio flux has a minimum (see Fig. 2). 
We constrain the significance of a possible radio flux decrease in 2008 by considering the non-detection in this epoch 
to have a flux between 0 and the $4\sigma$ upper limit. The possible flux decrease between the 1998 detection and the 2008 
non-detection has a significance 
between $3.3\sigma$ and $0.6\sigma$. The respective flux increase between 2008 and 2011 (the epoch with the maximum flux) 
has a significance between $6.0\sigma$ y $2.3\sigma$. These estimates do not include the absolute flux uncertainty of 
a few percent.
In the following subsections, we discuss several possible explanations for our result.
These possibilities do not necessarily exclude each other.

\begin{figure}[!ht]
\centering
\includegraphics[width=\columnwidth]{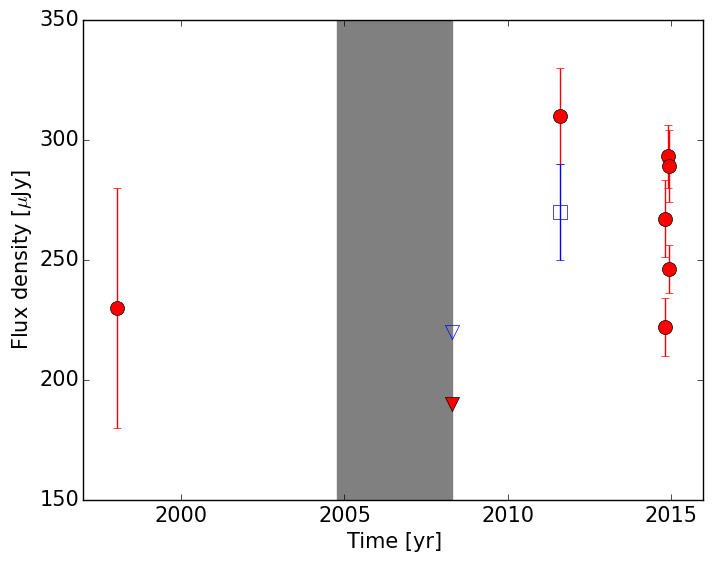}
\caption{\small Radio light curve of HOPS 383. The red dots and inverted triangles are X-band (8 to 10 GHz) measurements and 
upper limits, respectively. The blue symbols are in C-band (4 to 5 GHz). Error bars are $\pm 1\sigma$. 
See Table 1 for details of the observations. The vertical gray stripe marks the period of time that \cite{Safron15} 
estimated for the occurrence of the infrared burst. 
}
\label{fig2}
\end{figure}

\subsection{Protostellar accretion and ejection variations are not correlated in general.}

Accretion rates and jet/wind ejection rates
have been shown to be correlated within samples of the relatively evolved class II YSOs and FU Orionis 
objects \citep[e.g.,][]{Hartigan95,Calvet98}. However, the idea that in a given object accretion 
and ejection follow each other in time has only started to be tested. Contrary to the simplest expectations, 
it may be the case that they do not follow each other. \cite{Ellerbroek14} could not establish 
a relation between outflow and accretion variability in the 
Herbig Ae/Be star HD 163296, the former being measured from proper motions and radial velocities of the jet knots, 
whereas the latter was measured from near-infrared photometric and Br$\gamma$ variability. 
Similarly, \cite{CG14} monitored a sample of 19 embedded (class I) YSOs with near-IR spectroscopy and found that, 
on average, accretion tracers such as Br$\gamma$ are not correlated in time 
to wind tracers such as the H$_2$ and [Fe II] lines. 
The non-detection of a radio burst in HOPS 383 counterpart to the mid-IR burst is consistent 
with the above mentioned observational results. 

Numerical simulations that include the time-variable evolution of protostars can give insight on the possible 
time correlation of accretion and ejection. 
Some 3D MHD simulations \citep[e.g.,][]{Romanova09} have been performed with enough detail to follow the causal 
connection between protostellar accretion and jets+winds, but they follow the evolution of the system for timescales 
that are somewhat short compared to our observations and to FU-Ori like bursts in general.
Rather large, long bursts such as that observed in HOPS 383 are more likely the result of such processes as 
gravitational instabilities within the  
massive young circumstellar disk \citep[e.g.,][]{Voro15}, but these simulations do not yet resolve the physics of 
jet/wind production. 
Further work needs to be done to simulate what happens to the outflows in protostars with large accretion bursts.

\subsection{Protostellar accretion and ejection variations are not correlated in large bursts.}

Jets are launched through MHD processes that require the presence of an ordered magnetic field at 
the stellar-radii scale of the magnetosphere \citep{Shu94} and/or further out 
to a significant fraction of the disk \citep{KP00}. If in a bursting YSO the accretion 
rate becomes too high, it is possible that the configuration of the magnetosphere and the disk magnetic field 
become disrupted such that the launching of a collimated jet/wind is shut off \citep{Hartmann98}. 
This `jet-quenching' idea needs to be explored in detail in simulations.  

\subsection{The interpretation of the tracers.}

The final possibility that we consider is that either or both of the assumptions that the radio continuum traces 
free-free emission 
from a jet and that the mid-IR burst is due to an enhancement of accretion are incorrect. We argue that this is 
not the case. (Gyro)synchrotron emission 
from magnetospheres is known to contribute significantly in the more evolved class II and III YSOs 
\citep[e.g.,][]{Forbrich07}, but in class I and especially class 0 YSOs the radio emission is most likely dominated by the jet 
\citep[e.g.,][]{Dzib13,Liu14}. Furthermore, the spectral indices measured in the most sensitive epochs are consistent with 
free-free emission (see sections 2.3 and 2.4). Regarding the mid-IR burst, although the largest flux increase is at a
 wavelength of 24 
$\mu$m, \cite{Safron15} showed the occurrence of brightening from the near-IR to the submillimeter, 
making the case for a large ($\sim \times 35$) increase in bolometric -- and therefore accretion -- luminosity. 

A different instance of a misinterpretation of the observations would happen if  
the mid-IR burst and the radio emission arise from the different components 
of a binary system. Indeed, from Fig. 1, the radio emission can be 
interpreted as either a bipolar jet or a binary, with the features separated by $\approx 190$ AU 
\citep[0.45" at a distance of 414 pc,][]{Menten07}. 
However, even if the target is a binary, the lack of a large increase in the combined (unresolved) radio flux 
at any epoch remains to be explained.  

Finally, it is also possible that there was an ejection event traced by a radio jet (as suggested by the 
morphology in the epoch that resolved the source, see Fig. 1), but that the total flux density of the 
jet features did not have large variations. Based on a measurement of the proper motions in the 
lobes of the radio jet in IRAS 16293--2422, \cite{Pech10} inferred a recent bipolar ejection from this source. 
Interestingly, the total flux density at 8.5 GHz added over the lobes reported by those authors only increased 
by $\sim 40\%$ from 2003 to 2008. A large increase in the radio flux could rapidly fade if the ionized material 
recombines without further ionization. This could happen if some material is dense enough and gets shadowed 
from the ionizing source\footnote{The recombination timescale for gas with electron density 
$n_e \sim 10^6$ cm$^{-3}$ is only one month.} \citep[e.g.,][]{GM11}. 
If the possibility discussed here is correct, then time monitoring should be done at high angular resolution.

\section{Conclusions}

Our search in the VLA archive for a cm radio counterpart to the infrared burst recently reported in the class 0 
YSO HOPS 383 yielded a negative result. The lack of a counterpart in the radio jet to the 
accretion burst suggests that accretion and ejection variations 
do not follow each other in time, at least not for large ($\sim \times 35$) 
accretion enhancements and within short (up to a few years) periods of time. 

Our observations are consistent with recent reports, using different techniques, 
of a lack of temporal correlation between accretion and jet/wind 
tracers in class I YSOs \citep{CG14} and a Herbig Ae/Be star \citep{Ellerbroek14}. 
We discussed possible interpretations to these observations in the context of the available models 
of accretion/ejection from the boundary of a stellar magnetosphere with a disk \citep[e.g.,][]{Romanova09}. 
Time monitoring of more sources and more specific model predictions are needed to clarify the details of 
the connection between accretion and jet/wind ejection in YSOs.

\acknowledgements
This reasearch was done with the support of programs UNAM-DGAPA-PAPIIT IA101715 and 
UNAM-DGAPA-PAPIIT IA102815. R.G.M. thanks Jan Forbrich for comments on a draft of this Letter.
The authors thank the referee for a useful review.

\bibliographystyle{apj}
\bibliography{biblio}

\end{document}